\documentstyle[12pt]{article}
{   
\baselineskip\baselinestretch\baselineskip
\newcount\valueone  \valueone=\textheight
\newcount\valuetwo  \valuetwo=\baselineskip
\divide\valueone by \valuetwo
\advance\valueone by -2 
\message{number of lines per page = \the\valueone}
\global\textheight\valueone\baselineskip}
\begin{document}
\title{ Total Divergences in Hamiltonian Formalism\\
of Field Theory }
\author{ Vladimir O. Soloviev \thanks{e-mail: vosoloviev@mx.ihep.su}\\
{\it Institute for High Energy Physics}\\
{\it Protvino, Moscow Region, 142 284, Russia}}
\date{}
\maketitle


The field theory Hamiltonian formalism have been paid a lot of attention
in the two previous decades due to studying of integrable models.
Now the concepts of Poisson bracket and Hamiltonian operator are deeply
elaborated, see for example Refs.~1,2.

But from physicist's viewpoint the treatment based on free integration by
parts sometimes seems too restrictive. There are problems where surface
integrals  are nonzero and have physical
meaning. We can mention  characteristics of the gravitational field
in asymptotically flat space-times,$^3$ surface energy of a
fluid,$^4$ etc. As a result, in the cited works the concept of
Poisson bracket is exploited outside of the definition given in Refs.~1,2 and
in many other recent mathematical textbooks.

In this report it is proposed to generalize the definition of Poisson
brackets in order to treat integrals of divergences as normal
Hamiltonians which generate a kind of Hamiltonian equations on
the boundary. Nonlinear Schr\"{o}dinger equation is used as an
illustrative example.

We  use the local coordinate language and consider a domain $\Omega$
in  $R^n$ having a smooth boundary $\partial\Omega$.
The characteristic function of this domain is $\theta_{\Omega}
=\theta (P_{\Omega})$,
where equation $P_{\Omega}(x^1,...,x^n)=0$ defines the boundary.

\medskip
{\bf Definition 1}

An integral over  a finite domain $\Omega$
of  a function  of field variables $\phi^A(x), A=1,...,p$
and their partial derivatives $D_J\phi^A$ up to some finite order
$$
F=\int\limits_{\Omega}f(\phi_A(x),D_J\phi_A(x))d^nx
$$
is called a {\it local functional}.

All the functions $f$ and $\phi_A$ as well as their variations
are supposed to be infinitely smooth, i.e. $C^{\infty}(R^n)$.
We  use the multi-index notations $J=(j_1,...,j_n)$
$$
D_J={{\partial^{|J|}}\over{\partial^{j_1}x^1...{\partial^{j_n}x^n}}},
\qquad |J|=j_1+...+j_n.
$$
Binomial coefficients for multi-indices are
$$
{J \choose K}={j_1\choose k_1}...{j_n\choose k_n},
$$
We denote  as ${\cal  A}$ the space of local functionals . It is
important that this space includes functionals with integrands
depending on derivatives of arbitrary order. Otherwise the Poisson
bracket could go out of ${\cal  A}$.

{\bf Definition 2}

A bilinear operation $\{ \cdot,\cdot\}$   such that
for any $F,G,H\in~{\cal A}$

{\rm 1)} $\{F,G\}\in {\cal A}$;

{\rm 2)} $\{F,G\}=-\{G,F\}$;

{\rm 3)} $\{\{F,G\},H\}+\{\{H,F\},G\} +\{\{G,H\},F\}=0$;

is called the {\it new field theory Poisson bracket}.

The standard$^{1,2}$ definitions of local functional and Poisson bracket
differ only in adding words ``modulo divergences''.

It occurs possible to construct new expressions for
field theory Poisson brackets fulfilling the new definition,
by supposing locality and antisymmetry, i.e.
$$
\{ \phi_A(x),\phi_B(y)\}={1 \over 2}\sum_L \biggl( I_{AB}^L(x)D_L^{(x)}-
I_{BA}^L(y)D_L^{(y)}\biggr) \delta (x,y),
$$
and
these expressions are the same for all kinds of boundary conditions:
\begin{equation}
{1 \over 2}\sum_{A,B,P,Q}\int\limits_{\Omega} D_{P+Q}\biggl( E^P_A(f)\hat
I_{AB}E^Q_B(g) -E^P_A(g)\hat I_{AB}E^Q_B(f) \biggr)d^nx ,
\end{equation}
where
$$
\hat I_{AB}=\sum_NI^N_{AB}D_N,
$$
and coefficients $I^N_{AB}$  satisfy some standard conditions$^5$.
Higher Eulerian operators $E^J_A$ are defined in Ref.~1
through the formula of the full variation of a local functional
$$
\delta F=\sum_{A,J}\int\limits_{\Omega} D_J\biggl( E^J_A(f)
\delta\phi_A\biggr)d^nx.
$$
They can be explicitly expressed as
$$
E^J_A(f)=\sum_K (-1)^{|K|+|J|}{K\choose J}D_{K-J}{{\partial f}
\over{\partial\phi_A^{(K)}}}.
$$
Usual variational derivative (or Euler-Lagrange derivative)
is the Eulerian operator of zeroth order.
Let us mention that if $J$ is not included into $K$, then all
quantities having multi-index $(K-J)$ are zero. The sums over $J$
and $K$  are really finite because a local
functional can depend only on a finite number of derivatives.

The replacement of the old brackets by the new ones can be understood as
a transition from functional variations with fixed values of the field
on the boundary to free variations. Then a natural construction that we
call a full variational derivative appears in the place of the standard
Euler-Lagrange variational derivative. This new object cares information
not only on the integrand (up to constants), but also on the domain of
integration $\Omega$
$$
{{\delta F} \over {\delta\phi_A}}=E^0_A(\theta_{\Omega}f)=
\sum_J (-1)^{|J|}E^J_A(f)D_J\theta_{\Omega},
$$
where $\theta_{\Omega}$ is a characteristic function of the domain.

Inserting this variational derivative into the standard formula
$$
\{ F,G\} =\sum\limits_{A,B}\int\int {{\delta F} \over {\delta \phi_A(x)}}
\{ \phi_A(x),\phi_B(y) \} {{\delta G} \over {\delta \phi_B(y)}}d^nxd^ny,
$$
we should formally integrate by parts separately in $x$ and $y$ variables
to remove derivatives from the $\delta$-function.
Afterwards we take off one of the integrations naively with the help of
$\delta$-function and obtain an integral containing product of distributions.
By mystical way it occurs that by using a formal rule
$$
D_J\theta (P_{\Omega}) \times D_K\theta (P_{\Omega})=D_{J+K}\theta
(P_{\Omega})
$$
to transform integrand into the legal form and then integrating by parts
over infinite region we get a formula (1) that satisfies the new definition of
Poisson brackets given above.

Evidently to prove this statement we should
prove Jacobi identity and for wide classes of local brackets this is done
in Ref.~5.

Let us now consider a simple example of the Hamiltonian system and obtain
boundary equations with the help of the new brackets.
The nonlinear Schr\"{o}dinger
equation can be treated$^6$ as generated by the Hamiltonian
\begin{equation}
H=\frac{1}{2}\int({\cal H}+\bar{\cal H})dx,
\end{equation}
where
$$
{\cal H}=r' q' +kr^2q^2, \qquad \bar{\cal H}=\bar{r'}\bar{q'}+
k{\bar{r}}^2{\bar{q}}^2,
$$
and Poisson brackets are
$$
\{q(x),r(y)\}=-2i\delta (x,y), \qquad\{\bar{q}(x),\bar{r}(y)\}=2i\delta (x,y).
$$
To return to the standard notations we should put reality conditions
$$
\psi=q=\bar{r}, \qquad\bar\psi=r=\bar{q}.
$$
By considering Poisson brackets for integrals of the total spatial
derivatives of canonical variables $\phi_A=(q,r)$ with the Hamiltonian
we expect to obtain dynamical equations on the boundary in functional
form
$$
\frac{d}{dt}\int{\phi_A}' dx=
\{\int{\phi_A}' dx, H\}=\int D\bigl( \frac{1}{2}\sum\limits_BI_{AB}
{{\partial\cal H}\over{\partial\phi_B}}\bigr) dx.
$$
If we try to use Newton-Leibnitz formula then we get
$$
\dot{\phi_A}\bigg\vert_1^2=\frac{1}{2}\sum\limits_BI_{AB}
{{\partial\cal H}\over{\partial\phi_B}}\bigg\vert_1^2.
$$
This is different from the standard (internal) equations
$$
\dot{\phi_A}=\frac{1}{2}\sum\limits_BI_{AB}E^0_B(\cal H).
$$
For the given Hamiltonian the formal equations for boundary values are
$$
\dot q_b=-2ikr_b{q_b}^2, \qquad\dot r_b=2ik{r_b}^2q_b.
$$
These equations can be easily integrated and give elementary oscillations
at the ends:
$$
\psi_{b}(t)=\psi_{b}(0)exp(-2ik|\psi_{b}(0)|^2t).
$$
We can see that in this case the dynamics of boundary values is separated
from the internal dynamics. Of course, this situation is not general.

Let us now demonstrate that usual canonical transformations involving space
derivatives are not strictly canonical when divergence terms are taken into
account. In Ref.6  a transformation
$$
Q=\frac{1}{kr}, \qquad R=kr^2q -r'' +\frac{r'^2}{r},
$$
is exploited, which has an inverse
$$
r=\frac{1}{kQ}, \qquad q=kQ^2R-Q'' +\frac{Q'^2}{Q}.
$$
By taking $\phi_A=(Q,R)$ let us estimate
$$
\{ \phi_A(x),\phi_B(y)\} =\frac{1}{2}
(\hat I_{AB}(x)-\hat I_{BA}(y))\delta(x,y),
$$
and find that
\begin{displaymath}
\hat I_{AB}=2i
\left( \begin{array}{cc}
0 & -1\\
1 & -2D1/(kQ^2)D\\
\end{array} \right).
\end{displaymath}
In a recent paper$^8$ we introduce antisymmetric operators which
allow to simplify this expression.
So, noncanonicity of the transformation appears in the bracket
$$
\{ R(x),R(y)\} =\frac{-2i}{k}(D_x\frac{1}{Q^2(x)}D_x -D_y\frac{1}{Q^2(y)}
D_y)\delta(x,y).
$$
The Hamiltonian in the new variables will be given by Eq.(2) where
$$
{\cal H}=kQ^2R^2-Q'R' -2Q''R
+\frac{1}{k}\biggl[ 2\biggl( \frac{Q'}{Q}\biggr)^4 + \biggl(\frac{Q''}{Q}
\biggr)^2 -4\biggl(\frac{Q'^2Q''}{Q^3}\biggr)+\frac{Q'Q'''}{Q^2}\biggr] .
$$
We need to consider Hamiltonian equations for functionals
\begin{eqnarray}
F_1 & = & \int q'dx=\int \biggl(kQ^2R-Q''+\frac{Q'^2}{Q}\biggr)'dx,\nonumber\\
F_2 & = & \int r'dx=\int \biggl(\frac{1}{kQ}\biggr)'dx,\nonumber
\end{eqnarray}
which can be obtained by calculating the Poisson brackets
$$
\{ F,H\} = \{ F,H\} _{c} +\{ F,H\} _{nc},
$$
where
\begin{eqnarray}
\{ F,H\} _c & = &
-2i\sum\limits_{m,n}\int D_{m+n}\biggl( E^m_Q(F)E^n_R(H)-
E^m_R(F)E^n_Q(H)\biggr)dx,\\
\{ F,H\} _{nc} & = & -\frac{2i}{k}\sum\limits_{m,n}\int D_{m+n}\biggl(
E^m_R(F)D
\frac{1}{Q^2}DE^n_R(H)-
E^m_R(H)D\frac{1}{Q^2}DE^n_R(F)\biggr)dx.
\end{eqnarray}
Evidently, noncanonical term appears only in calculation of $\{ F_1,H\} $.

We first display here the higher Eulerian derivatives of $H$ and $F_1$:
\begin{displaymath}
\begin{array}{cccc}
E^0_R(H)=&kQ^2R-Q''/2,                          &E^0_R(F_1)=&0,\\
E^1_R(H)=&-Q'/2,                                &E^1_R(F_1)=&kQ^2,\\
E^0_Q(H)=&-R''/2+kQR^2,                         &E^0_Q(F_1)=&0,\\
E^1_Q(H)=&3/2R'+Q'^3/(kQ^4)-Q'Q''/(kQ^3),
                                       &E^1_Q(F_1)=&2kQR+({Q'/Q})^2-2Q''/Q,\\
E^2_Q(H)=&-R+Q'^2/(kQ^3)-Q''/(2kQ^2),         &E^2_Q(F_1)=&2Q'/Q,\\
E^3_Q(H)=&Q'/(2kQ^2),                           &E^3_Q(F_1)=&-1.
\end{array}
\end{displaymath}
Then we present the result of calculation for noncanonical term (4):
$$
-2i\int D\biggl( -Q'''' +3Q'''Q'/Q +3Q''^2/Q
-6Q''Q'^2/Q^2+2Q'^4/Q^3+2kQQ'R'+kQ^2R''\biggr)dx.
$$
Only taking this term into account we are getting for $\{ F_1,H\} $ a result
which is equivalent to the one obtained in  the old variables.

In one of our previous publications$^7$ we already have discussed this
problem in relation to Ashtekar's canonical transformation in General
Relativity. But there the standard definition of Poisson bracket was used
and so Jacobi identity was not granted to be fulfilled in general case.

\end{document}